# Tokenization of Real Estate Assets Using Blockchain

Shashank Joshi, SRM Institute of Science and Technology, India*

https://orcid.org/0000-0003-0072-4001

Arhan Choudhury, SRM Institute of Science and Technology, India

## ABSTRACT

Blockchain technology is one of the key technologies that have revolutionized various facets of society, such as the banking, healthcare, and other critical ecosystems. One area that can harness the usage of blockchain is the real estate sector. The most lucrative long-term investment is real estate, followed by gold, equities, mutual funds, and savings accounts. Nevertheless, it has administrative overheads such as lack of transparency, fraud, several intermediaries, title issues, paperwork, an increasing number of arbitrations, and the lack of liquidity. This paper proposes a framework that uses blockchain as an underlying technology. With the aid of blockchain and the suite of tools, it supports many of these problems that can be alleviated in the real estate investment ecosystem. These include smart contracts, immutable record management, tokenization, record tracking, and time-stamped storage. Tokenization of real estate lowers the entry barrier by fixing liquidity and interoperability and improving the interaction between various stakeholders.

## INTRODUCTION

Real estate is a one-of-a-kind and intricate asset type. In terms of asset base and transactional activity, the commercial real estate industry is an important global economic segment. Despite the size of the real estate investment sector, it has been dominated by a small group of corporate entities capable of making substantial, non-liquid investments. Unlike other asset classes, real estate has significant transaction costs, land-use rules, and other entry hurdles. These qualities of real estate have an impact on the market's overall efficiency. While there have been advancements in the flow of information and transactions among the various stakeholders, there are still many loopholes in the real estate ecosystem (Thota et al., 2019). Blockchain technology can address these significant loopholes in the real estate ecosystem and helps to digitize and increase the overall efficiency of the ecosystem. Blockchain is an immutable distributed ledger that keeps a record of all transactions and provenance of assets present in the blockchain network in the form of a block and a cryptographic code is used to link each block together in an orderly manner (Joshi, 2021). Furthermore, the desired result is achieved with the help of blockchain-based tokenization, which is a process based on blockchain technology that enables converting tangible and non-physical assets into blockchain tokens.

Real estate tokenization is a burgeoning industry, with an increasing number of people opting to go digital and have debt or equity tokenized. This process eliminates the centralized system that has existed in the past. The blockchain architecture is a preferable choice since it prevents unauthorized access and has other benefits such as immutability, better availability, and distributed nature. This technology transforms asset management in the form of transactions that are simple and automated in nature. Because real estate is typically an illiquid sector, and traditional techniques have lack of transparency, expensive processing fees, many intermediaries, and do not allow for fractional ownership; thus blockchain-based tokenization is the perfect solution for this ecosystem. Tokenization is the process of mapping and separating an asset into various tokens, each of which represents a certain portion of the value. The blockchain real estate token represents shares when one tokenizes a property i.e., a real estate asset. These shares can be used for a variety of purposes, including an equity interest, asset ownership, and dividend rights.

This article will examine how blockchain will transform the real estate sector by making transactions considerably faster, more secure, and transparent, as well as making investments safer and lowering transaction costs using blockchain as a framework for real estate asset tokenization, making it liquid, secure, and efficient. Extend the strategy to include an automated solution for token transfers and earnings distribution to investors.

The remainder of this paper is organized as follows. The next part presents some background or preliminaries for this work. Next, related work is presented in the same section. The paper then describes the existing real estate ecosystem and its flaws. This is followed by the proposed blockchain-based tokenization solution for the real estate ecosystem. The section after that discusses the impact of the proposed solution on the real estate ecosystem. Finally, the last section concludes this paper and indicates some future work.

## BACKGROUND AND RELATED WORKS

### Blockchain

The fundamental underlying technology used in our architecture is blockchain. It is an immutable distributed ledger that keeps a record of all transactions and provenance of assets present in the blockchain network in the form of a block and a cryptographic code is used to link each block together in an orderly manner (Joshi, 2021). If one block is to be added to the network, it needs to be approved and validated across the network by its peers. As blockchains are constantly growing, this adds to the security of the architecture. Blockchain fosters and maintains user confidence because it is a peer-to-peer system in which participants openly validate transactions using a consensus mechanism (Lunardi et al., 2019).

In a chain, each block contains data, a nonce, its own hash value, and the hash of the previous block. The nonce adds the cryptographic hash to each block as it is produced. The hash is now associated with the data, which has been signed. Each transaction is recorded with a time stamp. A genesis block is the first block in a chain. Furthermore, any device that retains blockchain copies is referred to as a node. Each participant is assigned a number that is unique to the user.

Blockchain networks broadly can be classified as private, public, or consortium-based in general as depicted Figure 1. A private blockchain is a variant of a blockchain network that restricts read and write access while also identifying the node responsible for validating and verifying transactions. As a result, transactions on private blockchain networks are less expensive and take less time to complete (Somin et al., 2018). Any user or node on the network can view and make transactions on a public blockchain. These blockchain networks are permissionless, which means that anyone can join them.

A consortium blockchain is a partially decentralized blockchain network in which a group of nodes is approved and responsible for the network's consensus.

## Ethereum Request for Comments (ERC)-1155

ERC-1115 is a multi-token standard that supports NFTs, semi-fungible and fungible tokens. It is the combination of ERC-20 and ERC-721 tokens, which are designed explicitly for fungible and non-fungible tokens, respectively.

The ERC-20 serves as a benchmark for fungible tokens, and in other words, which means that each one of them has the property of being precisely the same (in terms of type and value) as every other token. On the other hand, ERC-721 is an open standard for creating non-fungible, one-of-a-kind tokens on the Ethereum network.

ERC-1115 is better than the two standards as it handles both the tokens in a single smart contract at once, eliminating the complexity of handling two token standards and their interaction. It has emerged to be used in various fields, from gaming to real estate. Its new functionality is to create semi-fungible tokens capable of having both fungible token or NFT attributes; they start as fungible tokens and then become NFT, switching fungibility based on smart contracts.

The major differences between ERC1155, ERC721 and ERC20 tokens standards can be summarized as given in Table 1.

The multiple tokens help it transfer atomic swaps and escrows of various tokens on top of the ERC-1115 standard, with no need for approving contracts one by one as depicted in Figure 2.

Figure 1. Types of blockchain networks

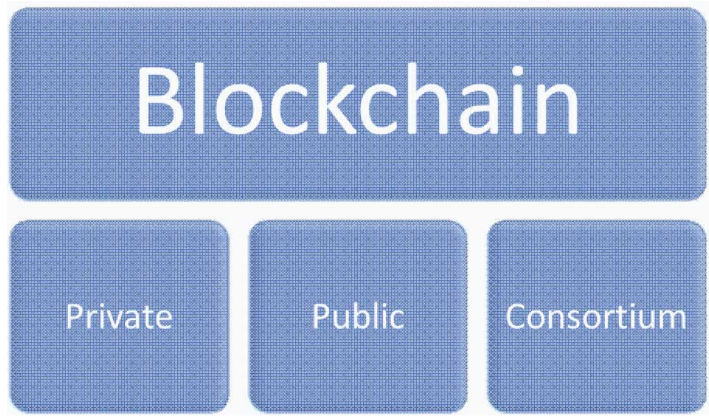

Table 1. Difference Between ERC-20, ERC-721 And ERC-1155

| Functionality | ERC-20 | ERC-720 | ERC-1155 |
| --- | --- | --- | --- |
| **Types Of Tokens** | Fungible or homogeneous tokens | Non-fungible or non-homogeneous tokens | Fungible and non-fungible tokens |
| **Properties** | Tokens properties are same | Tokens properties are different | Have quality of both ERC-20 and ERC-721 |
| **Divisibility And Interchangeability** | Tokens are losslessly interchangeable and divisible | Tokens are neither interchangeable nor divisible | Tokens are interchangeable and divisible to some extent |
| **Smart contracts** | Requires one common smart contract | Requires a unique smart contract for each token | Requires a single smart contract for infinite tokens |

**Figure 2. Swap in ERC-1155**

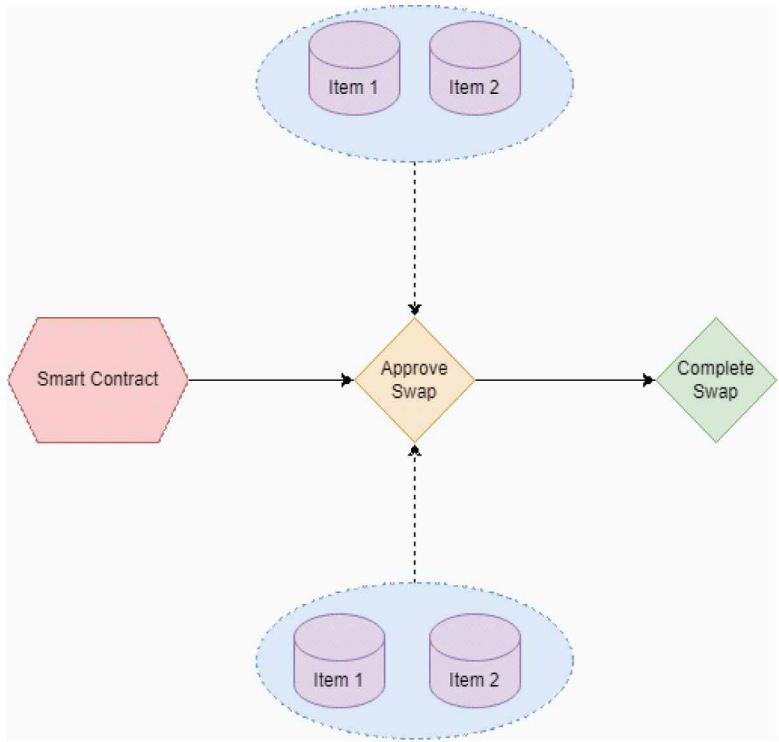

## Smart Contracts

As blockchain has varied features depending on the application, smart contracts are utilized to integrate the business logic with the blockchain network. Smart contracts are traceable and irreversible scripts that represent a real-world contract that may be enforced automatically in a decentralized system, minimizing the need for intermediaries (Christidis and Devetsikiotis, 2016). In a blockchain network, smart contracts are responsible for the execution of required business logic or operations required for the processing of any application, the immutability of created data, transparency, and auditability of completed processes or transactions.

## Ethereum Virtual Machine

Every Ethereum node has an Ethereum Virtual Machine, which operates as a decentralized computer and is in charge of executing bytecodes that constitute smart contracts (Buterin, 2013; Joshi, 2021). It is the environment in which smart contracts connected to the blockchain network run. Smart contracts enable peers in the network to change their state and return information about the current state by making requests. These requests are processed by blockchain nodes using the smart contract bytecode in their EVM, and the results are stored in the blockchain (Lunardi et al., 2019).

## InterPlanetary File System (IPFS)

Data may be stored and disseminated using the peer-to-peer distributed file system protocol known as IPFS (InterPlanetary File System). In a global namespace connecting all computer nodes or devices, IPFS uses content-based addressing to uniquely identify and locate any file.

## Related Work

Various studies have proposed strategies for resolving real estate sector difficulties with the help of blockchain-based architecture. Wright et al. (2015), for instance, look at the effect of blockchain on the function of economic and regulatory organizations in different important societal sectors. The blockchain-based architecture for real estate record-keeping was compared to the present system by Spielman et al. (2016), who also examined its advantages and drawbacks. The capabilities of blockchain and other distributed ledger technologies (DLT) are examined by Konashevych et al. (2020), along with the appropriateness of these technologies for diverse applications in real estate, property rights, and public records.

To better comprehend current research concerns, challenges, and future perspectives for blockchain deployment in e-Government, Batubara et al. (2018) have identified pertinent literature. In addition to proposing a design for safe paperless transactions for better asset management in a smart city, Karamitsos et al. (2018) also offered the methodology and functionality of each use case that were relevant to address the existing real estate concerns. According to Bhanushali et al. (2020), the adoption of blockchain-based architecture may be used to stop and prevent fraud during transactions between different real estate ecosystem parties. By using a Blockchain-based system to transform physical assets into immutable liquid Blockchain-based token assets, M. Nandi et al. (2020) provide a safe record-keeping technique that overcomes these issues. The advantages and disadvantages of utilizing a digital record to transfer properties on the blockchain as a safety net in case of litigation are examined by Laarabi et al. (2020).

## Current System and its Issues

Despite the significance of real estate in the country's economy, there are numerous flaws in the current system, including property information flow, lease agreements, liquidity, money transactions, and intermediary participation. Real estate assets can broadly be classified into residential, commercial, and industrial assets and all real estate assets fall into one of three categories. Buying an asset directly through a real estate broker is among the most traditional ways to invest in real estate. Real estate investments provide a number of advantages, including competitive risk-adjusted returns, high physical asset value, and attractive and consistent income flows in the form of rent and leasing fees.

Traditional real estate investment has a number of disadvantages as shown in Figure 3. To begin with, the initial investment required to purchase property is substantial. The majority of investors are unable to achieve this requirement and hence are unable to invest in real estate. Liquidity is also a problem for the system. Real estate investments are notoriously difficult to liquidate. The owner must also find suitable occupants in order to continue collecting rental money from the underlying property. It is impossible to sell a portion of an asset without selling the entire underlying item. Furthermore, the system typically involves a large number of mediators, including brokers, lawyers, and other professionals. The transaction fees involved in the real estate market are significant, and completing a real estate transaction takes a long time. All of these characteristics make the system inconvenient and unappealing to the average investor.

**Figure 3. Issues with current real estate sector**

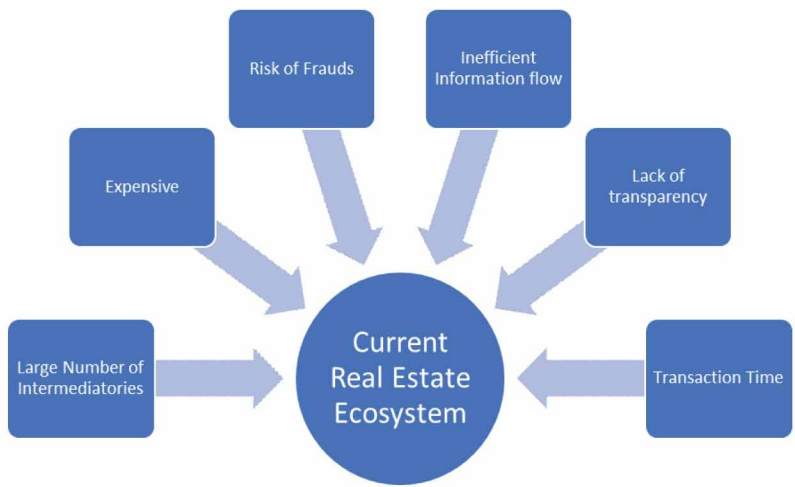

## PROPOSED SOLUTION AND ARCHITECTURE

In this paper, the authors consider the existing constraints and loopholes in the real estate sector such as High administrative overheads, ineffective information flow, high transaction time, lack of liquidity, etc. Based on these constraints, a blockchain-based tokenization system is proposed for optimizing the identified requirements and considerations. The roles and processes required for establishing real estate tokenization system smart contracts are explained and identified in the subsections that follow.

### Roles and Stakeholders

As represented in Figure 4, the proposed real estate tokenization system allows the participation of individuals or institutions in the following roles:

1. **Administrator/Validator:** It is in charge of overseeing and sustaining the blockchain system. Multiple trustworthy institutions and service points can be allocated to this role. The administrator is in charge of registering other stakeholders, validation of information used by the user to mint a token for an asset, and also has the authority to remove a stakeholder from the network under certain situations. It also assign permissioned nodes, thus enforcing the rules.
2. **Seller/Owner:** Sellers or Owners are the set of users in the network who looks into the following affordances:
   a. The seller is the one who wants to list a property based on its merits.
   b. It is responsible for the mapping of the asset in the network.
   c. The seller will interact with a buyer/tenant or realtor during the process of selling and renting.
   d. They will also need to participate in the legal processes that come associated with whatever action they want to perform (selling and renting).
   e. They set the price of a property (selling price or rent pricing).
3. **Buyer/Tenant:** Buyer or Tenant are the set of users in the network who looks into the following affordances:
   a. The buyer will choose and decide from a variety of options, be it renting or buying.
   b. They can be in the market for a variety of other things like mining, farming, business, etc.
   c. They will do a comparative study from all the options they get via the realtor or owner and decide based on the same.
   d. The buyer will also have to participate in the legal processes that come along with the affordance they trigger.
4. **Realtor:** The set institutes or individuals which look into the following affordances:
   a. The realtor is responsible for reaching out to their client network and finding potential buyers for a property.
   b. They are also responsible for selling the properties based on their various merits and qualities.
   c. Give the clients a comparative study of the variety of properties that suit their needs (The properties can be better assessed due to tokenization, for example - some properties might have good rental value so that the buyers can focus on the fractionalized rental asset on the other hand, if the property has reasonable value increments, then the buyers can focus on the fractionalized equity tokens).
  d. They are also responsible for handling the various affordances that are associated with the property:
     i. Buying - They will formalize the whole deal, i.e., the documentation, paperwork, title transfer, and legal affordances.
     ii. Renting - To formalize an agreement between a tenant and property owner.

## Interactions and Processes

In our proposed solution, the interactions between various stakeholders and processes of real estate sector are represented by the smart contract instantiated on the blockchain by the administrators. The following are the significant activities in the proposed system:

1. **Stakeholders Registration:** Administrators or permissioned nodes in the network handle the registration phase for various stakeholders of the network. This necessitates the verification of individuals and institutions in accordance with government norms, which necessitates a government verification service component. The creation of a wallet address accompanies each successful registration.
2. **Minting a Property Token:** The buyer will provide the administrator with details, specifications and documents (for example - property deed); the administrator will then verify these details from various institutional sources and certify them. Once these details are verified, the administrator will enter these details into a platform that will upload these into decentralized storage such as IPFS. These IPFS URIs will then be embedded into the base URI of the ERC-1155 property token and the NFT right tokens.
3. **Minting of fractional tokens:** The ERC1155 can mint both fungible and nonfungible tokens. These fungible tokens represent the fractional property right tokens, and the non-fungible will represent the property right token. The property right token and the fractional property right token are mapped to the real-world property using the custom URI in the base ERC1155 token. The fractional property rights tokens can be minted using the mint function in the property contract; this will fractionalize the property.
4. **Transferring fractional property tokens:** The fractional property token can be transferred into the buyer's wallet once it has been minted in the base ERC1155 property token, which represents the property.
5. **Upgradability:** This approach allows the smart contract to be upgraded in order to modify its code while retaining its balance, address, and state, allowing gradually adding more functions to the smart contract. Smart contracts, for example, can be upgraded if new rights, functionality, or other features are added.

**Figure 4. Roles and processes**

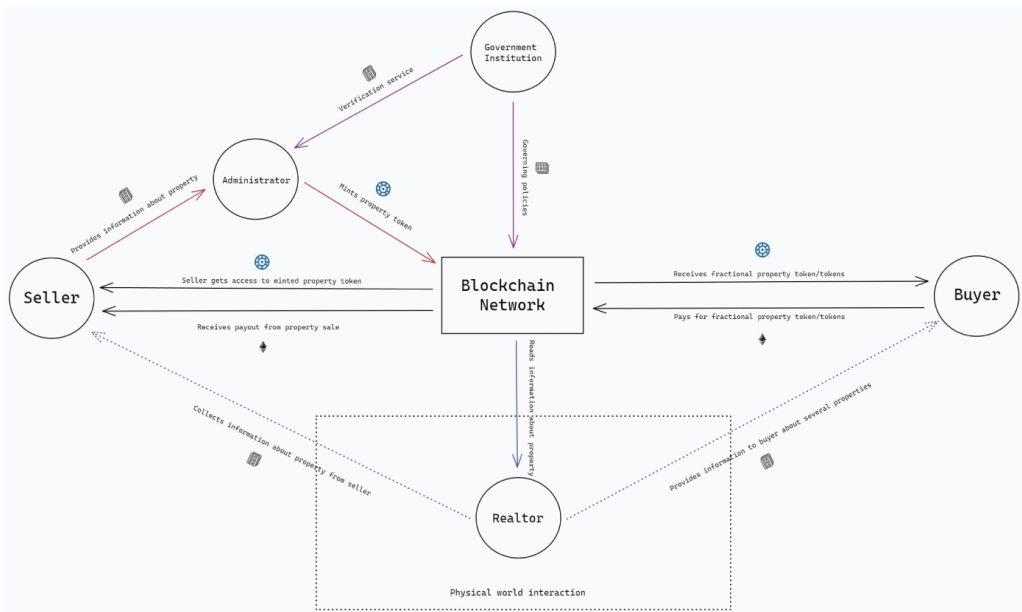

### Index of Functionalities

It's difficult to represent a property on a blockchain. When a physical asset is tokenized, it becomes part-owned. Let us assume a physical item with several owners is digitally represented to put things in perspective. This approach raises a slew of questions regarding who owns what. The ideal answer is to define the property before tokenizing it or to tokenize well-defined portions of the property. For the implementation of tokenization of real estate, the authors have come up with the following solution:

- ERC 1155 contracts are used as property contracts. The rights associated with each property are represented as non-fungible ERC1155, and the fractional rights are represented as fungible ERC1155 tokens.
- Two smart contracts are being used for this purpose:
  ◦ Property Contract - Used for representing individual properties.
  ◦ RealEstate Contract - Used for minting the property ERC1155 tokens.

This subsection will elaborate on the functionality of the proposed real estate tokenization system smart contract.

#### *Property Functionalities*

The property contract is an upgradable ERC1155 contract which is used to describe the property and is also used for tokenizing the property. The main reasons for using ERC1155 over a traditional ERC20 token are as follows:

1. Lower gas fees.
2. Better association of attributes of properties.

The ERC1155 is a multi-token standard, meaning it can contain fungible as well as non-fungible tokens. The individual property rights are represented as non-fungible tokens, and the tokenized property rights are represented as fungible tokens. The property right tokens will contain an IPFS hash of the required documents to prove its authenticity, and the fungible tokenized rights will be associated with these nonfungible tokens. The functions in the property contract are as follows:

- Function: initialize - It initializes the ERC1155 contract, the basic parameters are passed into this function by the factory contract, the parameters are described in the code. The uri parameter contains the base IPFS hash, it is set into the baseURI of the ERC1155. It contains all the information about the property.
- Function: getPropertyId - It returns the unique identifier associated with that particular property.

- Function: totalSupply - It returns the token supply of each token id (property right). In other words it returns the number of fractional tokens for the particular property right.
- Function: approvedProperty - It approves the property by comparing the parent hash in the merkle tree (Koo et al., 2018).
- Function: mintNFT - It is used to mint property right tokens.
- Function: mintBatchNFTs - It is used to mint a batch of or multiple property right tokens at once.
- Function: transferNFT - It allows the caller to transfer the NFT out of the ERC1155.
- Function: burnNFT - It is used to burn a property right token.
- Function: burnBatchNFTs - It is used to burn a batch of or multiple property right tokens at once.

### RealEstate Functionalities

The RealEstate contract is an upgradable contract with a base implementation, and further, for each property, proxies are generated. The functions used in the factory contract are described below:

- Function: deployPropertyContract - This function is responsible for deploying a proxy for each property. It emits an event that gives the deployed property address on the property token.
- Function: initialize - This function is responsible for initializing the factory contract; it takes in the implementation address, which is the base implementation of the property. It stores a variable value called logic address.
- Function: pause - This function enables the contract to be paused from minting; this is in the case of a bug or exploits being discovered.
- Function: unpause - This function enables the contract to continue functioning after the contract is once paused.
- Function: authorizeUpgrade - This function makes the whole factory upgradable; the caller can set a new implementation using this function.
- Function: getProxyLength - This function returns the length of the proxies (array).
- Array: proxies - It is an array of deployed contract proxy addresses. After each property contract is deployed, the proxy address is pushed into the proxies array.

Figure 5. Sample IPFS structure for the baseURI of ERC1155

```
{
  Name of Right: "Equity Right",
  Description: "This right depicts right of equity of so and so property",
  Documents: [
  {
      name: "Doc1",
      description: "something",
      link: "www.example-image.com"
  }
  ],
  ...
}
```

**Algorithm 1. Interface for Property Smart Contract**

```
interface Property {
  function initialize(address treasury,
          address upgrader,
          address admin,
          string memory uri,
          string memory _contractName,
          string memory _description
        ) external virtual;
  function getPropertyId() external view returns (uint256);
  function totalSupply(uint256 id) external view returns (uint256);
  function exists(uint256 id) external view returns (bool);
  function mintNFT(uint256 id, bytes memory data, uint256 price)
        external
        payable
        returns (uint256, uint256);
  function mintBatchNFTs(
        uint256[] memory ids,
        uint256[] memory amounts,
        bytes memory data,
        uint256[] memory prices )
        external payable returns (uint256[] memory,
      uint256[] memory);
  function burnNFT(address from, uint256 id, uint256 amount) external;
  function burnBatchNFTs(address from, uint256[] memory ids, uint256[]
  memory amounts) external;
  function
      transferNFT(
        address to,
        uint256 id,
        uint256 amount,
        bytes memory
      data
      ) external payable;
function approvedProperty(bytes memory _parentHash, address _propAddress)
external;
    }
```

## IMPACT OF THE PROPOSED MODEL ON REAL ESTATE SECTOR

A thorough examination shows that the proposed solution will maintain, improve, and emerge in real estate company plans and structures all over the world. Experts and individuals should contemplate this dependable reformist breakthrough in the sphere of real estate because it will make the process easier and less burdensome. Some of the major advantage of the proposed blockchain based architecture are:

- The use of tokens not only enables partial ownership, making it usually less expensive to participate in real estate projects, but it also prepares the way for the hiring of more financial professionals and the development of new economic sectors.
- The major advantage of tokenization is that, because all of its forms are stored on a blockchain, there are almost no third parties involved, which eliminates any possible frauds and expedites the process.

Algorithm 2. Interface for RealEstate Smart Contract

```
interface RealEstate {

function deployPropertyContract(
    address treasury,
    address upgrader,
    address admin,
    string memory uri,
    string memory _contractName,
    string memory _description
) external returns(address);
 function initialize() external;
function pause() external;
function unpause() external;
function _authorizeUpgrade(address newImplementation) external;
function getProxyLength() external view returns (uint256);
 }
```

- The proposed blockchain-based systems are invulnerable to debasement since they provide buyers and investors complete protection. Transparency, which is crucial in this industry, permeates every aspect of the suggested architecture.
- It provides unrivaled liquidity all across the globe. This could be due to the fact that it facilitates and secures cross-border token swaps or exchanges.
- Process automation with blockchain technologies ensures process efficiency for underlying industry activities and lowers expenses.

## CONCLUSION

In this paper, the focal objective was to resolve the loopholes in the real estate sector by leveraging the use of Blockchain technology. By removing the current flaws in the system, blockchain technology has the potential to enhance accountability, competency, and profitability in the real estate business. As blockchain technology advances, it is critical for the Real Estate industry to recognize its significance and examine its current business models, procedures, and strategies, resulting in a practical blockchain adoption plan that best suit this sector.

According to this study, the key to employing blockchain technology in real estate management is to create a digital mapping of a property in the blockchain. A smart token that contains or is related to data associated with the asset can be used to represent the property. Tokens are created by users to indicate their ownership rights. Users need trusted third parties to validate legal facts that they ordinarily cannot accomplish themselves, such as births, deaths, and notary acts. A trusted party is a broad concept here. The trusted third party generates a token that contains legal information about the user's token. As a result, the user's token is connected to the trusted third party's token. This ensures that the law will be followed. In combination with tokenization, smart contracts address concerns of security, liquidity, trust, and speed and enhance the token investments of real-world assets. Finally, it can be concluded that the suggested blockchain-based architecture is scalable, iterative, secure, and accessible and that it will address many of the difficulties that currently plague the real estate sector. As a next step, the authors intend to expand the proposed solution by introducing more functionalities for the user and to move toward the concept of digital ownership. Additionally, the authors also intend to test out this proposed solution on a variety of blockchain networks and consensus techniques.

# REFERENCES

Batubara, F. R., Ubacht, J., & Janssen, M. (2018). Challenges of blockchain technology adoption for e-government: a systematic literature review. *Proceedings of the 19th Annual International Conference on Digital Government Research: Governance in the Data Age*, 1–9. doi:10.1145/3209281.3209317

Bhanushali, D., Koul, A., Sharma, S., & Shaikh, B. (2020). Blockchain to prevent fraudulent activities: Buying and selling property using blockchain. In *2020 International Conference on Inventive Computation Technologies (ICICT)* (pp. 705–709). IEEE. doi:10.1109/ICICT48043.2020.9112478

Buterin, V. (2013). *A next generation smart contract & decentralized application platform (2013) whitepaper*. Ethereum Foundation.

Christidis, K., & Devetsikiotis, M. (2016). Blockchains and smart contracts for the internet of things. *IEEE Access: Practical Innovations, Open Solutions*, *4*, 2292–2303. doi:10.1109/ACCESS.2016.2566339

Joshi, S. (2021). *Feasibility of proof of authority as a consensus protocol model.* arXiv preprint arXiv:2109.02480.

Karamitsos, I., Papadaki, M., & Al Barghuthi, N. B. (2018). Design of the blockchain smart contract: A use case for real estate. *Journal of Information Security*, *9*(03), 177–190. doi:10.4236/jis.2018.93013

Konashevych, O. (2020). Constraints and benefits of the blockchain use for real estate and property rights. *Journal of Property, Planning and Environmental Law*.

Koo, D., Shin, Y., Yun, J., & Hur, J. (2018). Improving security and reliability in merkle tree-based online data authentication with leakage resilience. *Applied Sciences (Basel, Switzerland)*, *8*(12), 2532. doi:10.3390/app8122532

Laarabi, M., Maach, A., & Hafid, A. S. (2020). Smart contracts and over-enforcement: Analytical considerations on smart contracts as legal contracts. In *2020 1st International Conference on Innovative Research in Applied Science, Engineering and Technology (IRASET)* (pp. 1–6). IEEE.

Lunardi, R. C., Michelin, R. A., Neu, C. V., Nunes, H. C., Zorzo, A. F., & Kanhere, S. S. (2019a). Impact of consensus on appendable-block blockchain for iot. *Proceedings of the 16th EAI International Conference on Mobile and Ubiquitous Systems: Computing, Networking and Services*, 228–237. doi:10.1145/3360774.3360798

Lunardi, R. C., Nunes, H. C., Branco, V. S., Lipper, B. H., Neu, C. V., & Zorzo, A. F. (2019b). *Performance and cost evaluation of smart contracts in collaborative health care environments*. 10.20533/ICITST.WorldCIS.WCST.WCICSS.2019.0007

Nandi, M., Bhattacharjee, R. K., Jha, A., & Barbhuiya, F. A. (2020). A secured land registration framework on blockchain. In *2020 Third ISEA Conference on Security and Privacy (ISEA-ISAP)* (pp. 130–138). IEEE. doi:10.1109/ISEA-ISAP49340.2020.235011

Somin, S., Gordon, G., & Altshuler, Y. (2018). Network analysis of erc20 tokens trading on ethereum blockchain. In *International Conference on Complex Systems* (pp. 439–450). Springer. doi:10.1007/978-3-319-96661-8_45

Spielman, A. (2016). *Blockchain: Digitally rebuilding the real estate industry* [PhD thesis]. Massachusetts Institute of Technology.

Thota, S. S. (2019). Blockchain for real estate industry. *Scientific Review*, *5*(2), 53–56.

WrightA.De FilippiP. (2015). Decentralized blockchain technology and the rise of lex cryptographia. *Available at* SSRN 2580664. 10.2139/ssrn.2580664